\documentclass[conference]{IEEEtran}

\usepackage{amssymb}
\usepackage{times,epsfig}
\usepackage{moreverb}
\usepackage{amsthm}
\usepackage[cmex10]{amsmath} \interdisplaylinepenalty=2500
\ifCLASSINFOpdf
\else
\fi

\begin{document}
%
\title{Polar Coding for Achieving the Capacity of Marginal Channels in Nonbinary-Input Setting}

\author{\IEEEauthorblockN{Amirsina Torfi\IEEEauthorrefmark{1},
Sobhan Soleymani\IEEEauthorrefmark{1}, Seyed Mehdi Iranmanesh\IEEEauthorrefmark{1},
Hadi Kazemi\IEEEauthorrefmark{1},\\Rouzbeh A. Shirvani\IEEEauthorrefmark{2} and Vahid T. Vakili\IEEEauthorrefmark{3}}
\IEEEauthorblockA{\IEEEauthorrefmark{1}Department of Computer Science and Electrical Engineering,
West Virginia University, Morgantown, WV, USA\\ \IEEEauthorrefmark{2}Department of Electrical Engineering and Computer Science, Howard University, Washington D.C., USA\\
\IEEEauthorrefmark{3}Department of Electrical Engineering, Iran University of Science and Technology, Tehran, Iran\\
Email: \IEEEauthorrefmark{1}amirsina.torfi@gmail.com,
\IEEEauthorrefmark{1}ssoleyma@mix.wvu.edu,
\IEEEauthorrefmark{1}seiranmanesh@mix.wvu.edu,
\IEEEauthorrefmark{1}hadi.kazemi.azad@gmail.com,\\
\IEEEauthorrefmark{2}rouzbeh.asgharishir@bison.howard.edu,
\IEEEauthorrefmark{3}vakily@iust.ac.ir}}

\maketitle

\begin{abstract}
Achieving information-theoretic security using explicit coding scheme in which unlimited computational power for eavesdropper is assumed, is one of the main topics is security consideration. It is shown that polar codes are capacity achieving codes and have a low complexity in encoding and decoding. It has been proven that polar codes reach to secrecy capacity in the binary-input wiretap channels in symmetric settings for which the wiretapper's channel is degraded with respect to the main channel. The first task of this paper is to propose a coding scheme to achieve secrecy capacity in asymmetric nonbinary-input channels while keeping reliability and security conditions satisfied. Our assumption is that the wiretap channel is stochastically degraded with respect to the main channel and message distribution is unspecified. The main idea is to send information set over good channels for Bob and bad channels for Eve and send random symbols for channels that are good for both. In this scheme the frozen vector is defined over all possible choices using polar codes ensemble concept. We proved that there exists a frozen vector for which the coding scheme satisfies reliability and security conditions. It is further shown that uniform distribution of the message is the necessary condition for achieving secrecy capacity.
\end{abstract}


%
\IEEEpeerreviewmaketitle

\section{Introduction}


A notion of security named physical layer security defined by Wyner established an area of research in the world \cite{wyn75}. It uses the channel for its purpose by coding and other techniques in communication. Wyner's initial work assumes sender wants to send a confidential message to receiver without adversary being able to decode it. It was shown that if the channel from sender to receiver is statistically better than the channel from sender to adversary, secure and reliable communication is possible\cite{wyn75}.

\textit{M} is the \textit{k-bit} message that is supposed to be sent. Encoder maps \textit{M} to codeword \textit{X} and sends it on the channel. Bob receives sequence \textit{Y} on the main channel $W_m$ and Eve receives sequence \textit{Z} on the wiretap channel $W_w$. Finally decoder maps \textit{Y} to an estimate of message $\hat M $. The main purpose is to design a system for sending message reliably and secure on the channel when message length tends to infinity. Reliably is measured in terms of probability of error in rebuilding message. For a reliable system the following condition should be satisfied:
\begin{eqnarray}
\textbf{Reliability}: \mathop {\lim }\limits_{k \to \infty } \Pr (M \ne \hat M) = 0\
\end{eqnarray}

 Security is usually measured in terms of the normalized mutual information between the message \textit{M} and Eve's observations \textit{Z}. Encoding algorithms should satisfy the following to be called as secure:
 \begin{eqnarray}
\textbf{Weak Security}: \mathop {\lim }\limits_{k \to \infty } \frac{{I(M;Z)}}{k} = 0\
\end{eqnarray}

 Intuitively, (2) means that observing \textit{Z} does not provide much information about \textit{M}. Maurer discussed in \cite{mau94}, \cite{mau20} that the conventional notion of security (2) is a weak notion. Indeed, it is easy to construct examples where ${k^{1 - \varepsilon }}$ out of the \textit{k} message bits are disclosed to Eve while still satisfying (2). So Maurer defined another condition in \cite{mau94}:
 \begin{eqnarray}
\textbf{Strong Security}: \mathop {\lim }\limits_{k \to \infty } I(M;Z) = 0
\end{eqnarray}

Notice that both security conditions (2) and (3) are information-theoretic not computational: there is not any limitation for computational power of adversary, and security does not depend on computational complexity of algorithm. Basically the leakage of the information must be prevented and it is unlike the other works like \cite{vapnik15}, \cite{moti16} and \cite{lampert13} in which the marginal information is leveraged as auxiliary information for reasoning.

\subsection{Assumptions and Settings}
Our coding scheme is based on polar coding presented in\cite{ari09}. Polar codes achieve the capacity of binary-input DMCs. It is shown in \cite{ari09} that for sufficiently large block length the channel turn into a perfect channel or a complete noisy one. Noiseless channels are called good channels and the other ones are called bad channels.

Main idea of our construction is as following: Random symbols are transmitted over those channels that are good for both Eve and Bob. Information symbols are transmitted over those channels that are good for Bob but bad for Eve, and a fixed vector is transmitted over those channels that are bad for both Bob and Eve which is defined over all possible choices using polar coding ensemble. We prove that there exist a sequence of frozen symbols that our coding scheme satisfies the reliability and security condition. In our proposed coding scheme we consider asymmetric settings and wiretap channel $W_w$ is degraded with respect to $W_m$ and both have nonbinary-input alphabet.

\subsection{Backgrounds and Related Works}
Wyner considered a condition in which both main and wiretap channels are discrete memoryless channels and wiretap channel, is degraded with respect to main channel. He proved that for rates lower than a constant number he called secrecy capacity, reliable and secure transfer of data is possible \cite{wyn75}.\\
After Wyner, some works fulfilled in this area like \cite{che78} and \cite{csi78}. In some of them like \cite{che09}, \cite{oza84} a constraint is assumed on the computational power of eavesdropper.

 A large number of works in information-theoretic security, prove the existence of a code sequences which using them leads to achieving secrecy capacity, but a few number of them design explicit encoding-decoding algorithms. Explicit algorithms was presented in \cite{sur10} and \cite{tha07} and led to achieving secrecy capacity. In \cite{oza84} a situation is studied that assume a computational power constraint on adversary.
 
  There are other works similar to ours like \cite{mah11},\cite{hof10} and \cite{koy12} but main differences should be mentioned. \cite{hof10} and \cite{mah11} considered binary-input channels for symmetric setting but our contributions is for asymmetric nonbinary-input channels. Moreover the focus of \cite{koy12} is on utilizing privacy amplification protocol and information distillation for key agreement but we only consider pure information-theoretic security. In \cite{hof10} the distribution of message is assumed to be uniform But we don't consider any specific distribution on the message which is a fair assumption on message \textit{M}.

\subsection*{Organization}
In Section II, relevant concepts related to wiretap channels, in order to show an expression for the secrecy capacity  is represented in the setting where $W_m$ and $W_w$ are DMC and $W_w$ is degraded with respect to $W_m$. Also the notion of symmetric channels and secrecy capacity has been defined.
Section III is devoted to polar codes and important theorems that are necessary for our proofs.
We present the proposed coding scheme in Section IV and proofs show the security and reliability of proposed scheme and also the code rate approached to secrecy capacity while the message distribution is uniform. 
We conclude the paper in Section VI with a brief discussion of further results.
\subsection*{Notation definition}
 We denote random variables (RVs) by upper case letters, such as \textit{X}, \textit{Y} and their sample values by the corresponding lower case letters, such as $x,y$ and calligraphic font shows the alphabet set of related random variable. For example, ${\cal X}$ shows the alphabet set of \textit{X} and $|{\cal X}|$ demonstrates the alphabet size. $P_X$ is the distribution of \textit{X}. If $f(x)$ and $g(x)$ are defined on a subset of real numbers then we write 
$f(x) = O(g(x))$ if for large $x$ there exists a constant number \textit{M} for which the inequality 
$f(x) \le M(g(x))$ holds. Notation $a_1^N$ is used instead of row vector $({a_1},{a_2},...,{a_N})$ and notation ${a_A}$ is used for representing the sub vector $\left( {{a_i}:{\rm{ }}i \in A} \right)$.
$C_W$ is the capacity of the channel W and defined as :
\begin{eqnarray}
C = \mathop {\max }\limits_{{P_X}} I(X;Y)
\end{eqnarray}
$I_W$ is the symmetric capacity of the channel W and for general channels defined as below:
\begin{eqnarray}
I(W) = \sum\limits_{y \in {\cal Y}} {\sum\limits_{x \in {\cal X}} {\frac{1}{{|{\cal X}|}}} } W(y|x)\log \frac{{W(y|x)}}{{{\textstyle{1 \over {|{\cal X}|}}}\sum\limits_{x' \in {\cal X}} {W(y|x')} }} 
\end{eqnarray}

This is the mutual information when the input distribution
$P_X$ is uniform.

$[N]$ stands for $\{ 1,...,N\}$. $log(.)$ is based on 2 in the rest of the paper.
\vspace{-3mm}
\section{Secrecy Capacity}
In this section we review the prior works in \cite{csi78} and \cite{s77}. Our discussion is about nonbinary discrete memoryless channels. Such channel has the transition probability of $W(y|x)$ and ${\cal X}$,${\cal Y}$ as input and output alphabets of channel. 
 
Assume there exists a channel $W_b$ with input alphabet ${\cal Y}$ that holds following equation:
\begin{eqnarray}
{W_w}(z|x) = \sum\limits_{y' \in {\cal Y}} {{W_m}(y'|x)} {W_b}(z|y'){\rm{   }}\forall x,z 
\end{eqnarray}

In this case, the wiretapper's channel is stochastically degraded with respect to the main channel. According to \cite{csi78} the secrecy capacity is:
\begin{eqnarray}
{C_s} = \mathop {\max }\limits_{{P_X}} \{ I(X;Y) - I(X;Z)\}
\end{eqnarray}
\vspace{-3mm}

According to the concept of symmetric capacity and equation (7), by assuming uniform input distribution $P_X$, the capacity-equivocation region is given by:
\begin{eqnarray}
{R_e} \le R \le {I_{{W_m}}},0 \le {R_e} \le {I_{{W_m}}} - {I_{{W_w}}}
\end{eqnarray}
 In this case the secrecy capacity define as:
\begin{eqnarray}
{C_s} = {I_{{W_m}}} - {I_{{W_w}}}
\end{eqnarray}
\vspace{-5mm}
\vspace{-5mm}
\section{Polar Coding}
In this section important notion of polar coding is defined which is used in our designs and proofs.
\subsection{Primitive Definitions}
Consider we have a binary-input DMC which its transition probabilities are $W(y|x)$ . \textit{N} transition over the channel is shown by ${W^N}(y_1^N|x_1^N)$.
 The Bhattacharyya for nonbinary input DMCs with alphabet size q is defined as one of the two following forms\cite{sas09}:
    \begin{align}
    Z(W) = \frac{1}{{1 - q}}\sum\limits_{x,x' \in {\cal X}:x \ne x'} {\sum\limits_y {\sqrt {W(x,y)W(x',y)} } } 
\end{align}     

  \begin{align}
    Z(W) = \frac{1}{{1 - q}}\sum\limits_{x,x' \in {\cal X}:x \ne x'} {\sum\limits_y {\sqrt {W(y|x)W(y|x')p(x)p(x')} } } 
\end{align}  

If uniform input distribution is assumed we have:
\begin{align}
    Z(W) = \frac{1}{{q(1 - q)}}\sum\limits_{x,x' \in {\cal X}:x \ne x'} {\sum\limits_y {\sqrt {W(y|x)W(y|x')} } } 
\end{align}  

 Channel polarization has two phases: channel combining and channel splitting. $u_1^N$ is the transformed vector. The combined channel is formed by :
\begin{eqnarray}
{W_N}(y_1^N|u_1^N)  \equiv  {W^N}(y_1^N|u_1^N{G_N})
\end{eqnarray}
$G_N$ is the generator matrix of the polar code.
In the channel splitting phase, channels formed by:
\begin{eqnarray}
W_N^{(i)}(y_1^N,u_1^{i - 1}|{u_i}) \equiv \frac{1}{{{2^{N - 1}}}}\sum\limits_{u_{i + 1}^N \in {X^{N - i}}} {{W_N}(y_1^N|u_1^N)} {\rm{ }}
\end{eqnarray}
The idea of polar coding is utilizing the polarization phenomenon and transmitting data using the channels for which $I(W_N^{(i)})$ is approximately 1.

 In \cite{ari09} the notation ${\cal P}(N,K,{\cal A},{u_{{{\cal A}^c}}})$ is defined for polar code. ${u_{{{\cal A}^c}}}$ is the frozen vector . Information set (${\cal A}$) is chosen such that it satisfies two conditions :first $|{\cal A}| = K$ and second $Z(W_N^{(i)}) \le Z(W_N^{(j)})$ for all $i \in {\cal A},j \in {{\cal A}^c}$. The decode has the knowledge of frozen vector. Decoder(Successive cancellation decoder) derive an estimation of the input :for frozen indexes put ${u_{{A^c}}} = {{\hat u}_{{A^c}}}$ and For other indexes which satisfying $i \in {\cal A}$ it puts ${{\hat u}_i} = 0$ if ${\mathop W\nolimits_N^{(i)} (y_1^N,\hat u_1^{i - 1}|1) \le \mathop W\nolimits_N^{(i)} (y_1^N,\hat u_1^{i - 1}|0)}$ and otherwise ${{\hat u}_i} = 1$.
 
  Recently, the results of \cite{ari09} has extended to general DMCs in \cite{sas09}. These results show polar coding usage for achieving the symmetric capacity of the DMC channels with alphabet of size q given by (2) in which $|{\cal X}| = q$.
  
  Channel polarization, from the early stage proposed for binary-input settings,and now has been generalized to arbitrary DMCs. It is proved that when the size of input alphabet is prime, a similar method to binary-input case leads to polarizing the channels. This method extended to channels of not-prime input sizes by splitting these channels to a subset of channels that have prime input alphabet size. The extending of polarization for channels with input alphabet of arbitrary sizes leads to polar coding for achieving the real capacity of arbitrary channels\cite{sas09}. 
 
 In this paper we skip further details of polar coding for nonbinary input channels and use the results. For more details, see \cite{ari09} and \cite{sas09}.
  
For the rest of the paper Z(W) means equation (12).
\subsection{Polar Coding Ensemble}
Now, polar code notation is represented which is implemented for defining polar coding ensemble and is used for the rest of the paper.

\textbf{Definition 1} (\textit{Polar Coding\cite{ari09}}):
Polar code ${{\cal P}}(N,{{\cal A}},{u_{{\cal F}}})$ for every ${{\cal A}} \subseteq \{ 1,...,N\} $ and ${u_{{\cal F}}} \in {{{\cal X}}^{|{{\cal F}}|}}$ is a linear code according to the following notation:
\begin{eqnarray}
{{\cal P}}(N,{{\cal A}},{u_{{\cal F}}}) = \{ x_1^N = u_1^N{G_N}:{u_{{{{\cal F}}^c}}} \in {{{\cal X}}^{|{{{\cal F}}^c}|}}\}
\end{eqnarray}

${{\cal F}}$ is frozen set and its indexes are called forzen indexes. Also ${{\cal A}}$ is information set and its indexes are called information indexes.

\textbf{Definition 2} (\textit{Polar Coding Ensemble\cite{kor09}}):Polar code Ensemble ${{\cal P}}(N,{{\cal A}})$ for every ${{\cal A}} \subseteq \{ 1,...,N\} $ represents the Ensemble below:
\begin{eqnarray}
{{\cal P}}(N,{{\cal A}}) = \{ {{\cal P}}(N,{{\cal A}},{u_{{\cal F}}}):\forall {u_{{\cal F}}} \in {{{\cal X}}^{|{{\cal F}}|}}\} 
\end{eqnarray}
${P_{B,e}}({{\cal A}},{u_{{\cal F}}})$ represents the block error probability ${{\cal P}}(N,{{\cal A}},{u_{{\cal F}}})$ with uniform distribution assumption on all codewords. ${P_{B,e}}({{\cal A}})$ is average block error probability of ensemble ${{\cal P}}(N,{{\cal A}})$ means averaging ${P_{B,e}}({{\cal A}},{u_{{\cal F}}})$ on all possible choices of ${u_{{\cal F}}} \in {{{\cal X}}^{|{{\cal F}}|}}$ with equal probability.

\textbf{Lemma 1} (\textit{Averaged Block Error Probability Upper Bound}\cite{kor09}): for a B-DMC W and information set ${{\cal A}}$, averaged error probability of block (over all possible choices of frozen vectors) can be bounded as follows:
\begin{eqnarray}
{P_{B,e}}({{\cal A}}) \le \sum\limits_{i \in {{\cal A}}} {Z(W_N^{(i)})}  
\end{eqnarray}

\subsection{Rate of Polarization}
following theorem shows the concept of polarization :\\
\textbf{Theorem 1} (\textit{Rate of Convergence\cite{ari09}}):for any binary discrete memoryless channel W, for $N = {2^n}$ and $\delta  \in (0,1)$ :
\begin{eqnarray}
\mathop {\lim }\limits_{N \to \infty } \frac{{|i \in [N] :I(W_N^{(i)}) \in (1 - \delta ,1)|}}{N} = I(W) 
\end{eqnarray}
\vspace{-5mm}
\begin{eqnarray}
\mathop {\lim }\limits_{N \to \infty } \frac{{|i \in [N] :I(W_N^{(i)}) \in (0,\delta )|}}{N} = 1 - I(W)
\end{eqnarray}

\textbf{Theorem 2} (\textit{\cite{sas09}}): For any DMC W in which $ R < I(W)$, parameter $\beta \in (0,1/2)$ is considered to be fixed. Averaged Block error probability of polar coding satisfy the following equality:
\begin{eqnarray}
{P_{B,e}}({{\cal A}}) \le (q-1)*{2^{ - {N^\beta }}}
\end{eqnarray}
In which q is the input alphabet size.
Now we propose a lemma which is an extention of lemma 4.7 from \cite{kor09} that used for realizing good channels and bad channels from each other.

\textbf{Lemma 2}:
if $W:{\cal X} \to {\cal Y}$ and ${W_d}:{\cal X} \to {{\cal Y}_d}$ are two DMC W with nonbinary input alphabet size and $W_d$ is degraded with respect to $W$ then there exists a channel like ${W_b}:{\cal Y} \to {{\cal Y}_d}$ that ${W_d}({y_d}|x) = \sum\limits_{y \in {\cal Y}} {W(y|x){W_b}({y_d}|y)} $. In this condition ${W_d}_N^{(i)}$ is degraded with respect to ${W}_N^{(i)}$ and $Z({W_d}_N^{(i)}) \ge Z(W_N^{(i)})$.
\begin{proof}
now we should prove that inequality $Z({W_d}_N^{(i)}) \ge Z(W_N^{(i)})$ holds:\\
\begin{flalign}
&Z({W_d}_N^{(i)}) = \frac{1}{{1 - q}}\sum\limits_{x,x' \in {\cal X}:x \ne x'} \nonumber\\&{\sum\limits_{{y_d} \in {{\cal Y}_d}} {\sqrt {{W_d}_N^{(i)}({y_d}|x){W_d}_N^{(i)}({y_d}|x')p(x)p(x')} } }\nonumber&
\end{flalign}
\vspace{-5mm}
\begin{flalign}
& = \frac{1}{{1 - q}}\sum\limits_{x,x' \in {\cal X}:x \ne x'} {\sqrt {p(x)p(x')}} \nonumber\\ & \times \sum\limits_{{y_d} \in {{\cal Y}_d}} {\sqrt {{W_d}_N^{(i)}({y_d}|x){W_d}_N^{(i)}({y_d}|x')} } \nonumber &
\end{flalign}
\vspace{-5mm}
\begin{flalign}
& = \frac{1}{{1 - q}}\sum\limits_{x,x' \in {\cal X}:x \ne x'} {\sqrt {p(x)p(x')} }  \times \nonumber\\ & \sum\limits_{{y_d} \in {{\cal Y}_d}} {\sqrt {\sum\limits_{y \in {\cal Y}} {W_N^{(i)}(y|x){W_b}_N^{(i)}({y_d}|y)} \sum\limits_{y \in {\cal Y}} {W_N^{(i)}(y|x'){W_b}_N^{(i)}({y_d}|y)} } } \nonumber &
\end{flalign}
\vspace{-5mm}
\begin{flalign}
&\mathop  \ge \limits^{(a)} \frac{1}{{1 - q}}\sum\limits_{x,x' \in {\cal X}:x \ne x'} {\sqrt {p(x)p(x')} } \nonumber\\& \times \sum\limits_{{y_d} \in {{\cal Y}_d}} {\sum\limits_{y \in {\cal Y}} {\sqrt {W_N^{(i)}(y|x){W_b}_N^{(i)}({y_d}|y)W_N^{(i)}(y|x'){W_b}_N^{(i)}({y_d}|y)} } } \nonumber&
\end{flalign}
\vspace{-5mm}
\begin{flalign}
& = \frac{1}{{1 - q}}\sum\limits_{x,x' \in {\cal X}:x \ne x'} {\sqrt {p(x)p(x')}} \nonumber\\ & \times \sum\limits_{{y_d} \in {{\cal Y}_d}} {\sum\limits_{y \in {\cal Y}} {{W_b}_N^{(i)}({y_d}|y)\sqrt {W_N^{(i)}(y|x)W_N^{(i)}(y|x')} }  } \nonumber &
\end{flalign}
\vspace{-5mm}
\begin{flalign}
& = \frac{1}{{1 - q}}\sum\limits_{x,x' \in {\cal X}:x \ne x'} {\sqrt {p(x)p(x')} }\nonumber\\ & \times  \sum\limits_{y \in {\cal Y}} {\sqrt {W_N^{(i)}(y|x)W_N^{(i)}(y|x')} \sum\limits_{{y_d} \in {{\cal Y}_d}} {{W_b}_N^{(i)}({y_d}|y)} }  = Z(W_N^{(i)})&
\end{flalign}

(a) follows from Cauchy-Schwartz inequality.
\end{proof}
This lemma implies that with assumption of degradation of wiretap channel with respect to main channel if a channel is good to Eve it is good for Bob, conversely if a channel is bad for Bob, it is Bad for Eve too.
\vspace{-5mm}
\section{Proposed Coding Scheme}
In this section we represent a coding scheme and prove its security and reliability . Also we show that it achieves the rate of secrecy capacity.
\subsection{Secret Codebook}

Assume that Alice has a confidential message \textit{M} which is to be transmitted to Bob and should be hidden from the Eve.
Reliable and secure communication is possible, if the following conditions for any given $\varepsilon  > 0$ be satisfied:
\begin{eqnarray}
Reliability :{P_{B,e}}({{{\cal A}}_m}) \le \varepsilon 
\end{eqnarray}
\vspace{-5mm}
\begin{eqnarray}
Security :\frac{1}{N} I(M;Z_1^N) \le \varepsilon 
\end{eqnarray}
Wiretap channel is degraded with respect to the main channel. We have the following Markov chain:
\begin{eqnarray}
U \to X \to (Y,Z) 
\end{eqnarray}
for sufficiently large N and $0 < \beta  < 1/2$ following sets are defined:
\begin{eqnarray}
{{{\cal A}}_m} = \{ i \in [N] :Z({W_m}_N^{(i)}) \le \frac{q-1}{N}{2^{ - {2^{n\beta} }}}\} 
\end{eqnarray}
\vspace{-5mm}
\begin{eqnarray}
{{{\cal F}}_m} = \{ i \in [N] :Z({W_w}_N^{(i)}) \ge \frac{q-1}{N}{2^{ - {2^{n\beta} }}}\}  
\end{eqnarray}
According to the defined sets, ${{{\cal A}}_m}$ is the good channel indexes for the main channel and ${{{\cal F}}_m}$ is the bad channel indexes for the wiretap channel.
According to polar coding definition and lemma 2 it is concluded that ${{\cal F}_m} \subseteq {{\cal F}_w},{{\cal A}_w} \subseteq {{\cal A}_m}$. The main task is to form $u_1^N$ based on defined indexes sets.
$u_1^N$ is the vector that multiplied by generator matrix and forms transmitted codeword. Fig.1 demonstrates the relation between defined indexes.
 \begin{figure}[h!]
  \centering
           \includegraphics[width=0.4\textwidth]{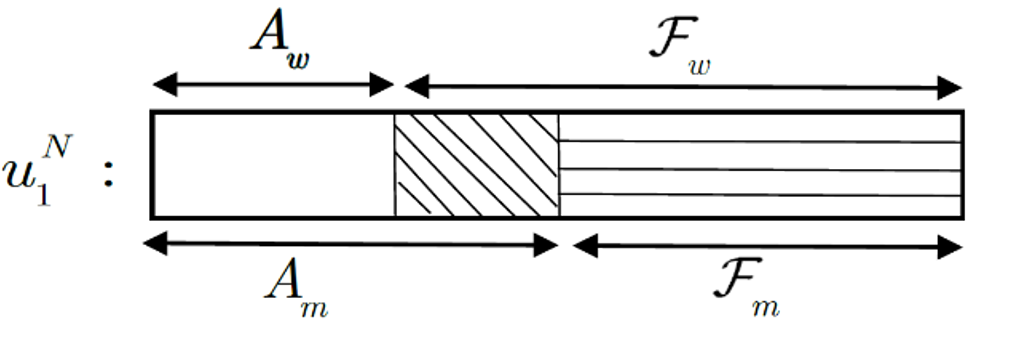}
           \centering
  \caption{the relation between main channel and wiretap channel indexes}
\end{figure}
In the following part the coding scheme has been proposed. According to definition 2 we define our proposed coding scheme by defining frozen symbols over all possible choices.
\subsection{Encoding Algorithm}
Secret message is mapped on the vector $V_m$ and random vector $V_r$ is generated with uniformly random distribution. Now the vector $u_1^N$ is formed as below:
\begin{enumerate}
\item Information symbols are sent over the indexes which are good for the main channel and bad for the wiretap channel. This concept could be shown by ${u_{{{\cal A}_m}\backslash {{\cal A}_w}}} = {u_{{{\cal F}_w}\backslash {{\cal F}_m}}} = {u_I}$
message length is k and $|{V_m}| = |{{\cal A}_m}| - |{{\cal A}_w}| = k$.
\item We send random symbols over indexes that belong to good channel for both main and wiretap channels as ${u_{{{\cal A}_w}}} = {u_R}$.
\item Over bad channels for both equal ${u_{{{\cal F}_m}}}$ we send a frozen vector assumed to be chosen from all possible choices ${u_{{{\cal F}_m}}} \in {{{\cal X}}^{|{{{\cal F}}_m}|}}$ and given to decoder of Bob and Eve. Polar coding ensemble is used, it meants coding scheme is built over $\forall {u_{{{\cal F}_m}}} \in {{{\cal X}}^{|{{{\cal F}}_m}|}}$. 

\end{enumerate}
Encoding is defined as the following function:
\begin{align}
&encod:{\{ 0,...,q - 1\} ^k} \times {\{ 0,...,q - 1\} ^r} \to {\{ 0,...,q - 1\} ^N}&
\end{align}
Input contains a secret message $m \in {\{ 0,1,...,q - 1\} ^k}$ and a random vector ${u_R} \in {\{ 0,1,...,q - 1\} ^r}$. There is no assumption on the message distribution but $u_R$ assumed to be choosen randomly from ${\{ 0,1,...,q - 1\} ^r}$ with uniform distribution.
\subsection{Decoding}
Decoding is divided into two subsection. Reliability satisfaction and coding rate.
\subsubsection{Reliability}
Both $V_m$ and $V_r$ has been defined over good indexes of main channel, thus according to theorem 3, both of them could be decoded using SC decoding with probability of error as ${P_{B,e}}({{{\cal A}}_m}) = O((q-1)*{2^{ - {N^\beta }}})$ \cite{sas09}(averaging over all possible choices of ${u_{{{\cal F}_m}}}$). So the reliability condition is satisfied.
\subsection{Security Proof}
In this part we prove the security of the coding scheme.
Since the scheme is formed over all possible choices of frozen symbols(polar coding ensemble), mutual information between message and Eve evaluated using one random chosen vector ${u_{{{\cal F}_m}}}$ over the whole set ${u_{{{\cal F}_m}}} \in {{{\cal X}}^{|{{{\cal F}}_m}|}}$. After choosing ${u_{{{\cal F}_m}}}$ we fix it and ultimately we should prove that there exists such ${u_{{{\cal F}_m}}}$ for our purpose. The decoding error probability of Eve has been evaluated over the ensemble in average sense. Now we have the following equations:
\begin{align}
&I(M;Z|{U_F}) = I(U;Z) - I({U_R};Z|{U_F},{U_I})\nonumber\\
&\mathop  = \limits^{(a)} I(U;Z) - H({U_R}|{U_F},{U_I}) + H({U_R}|{U_I},{U_F},Z)\nonumber\\
&\mathop  = \limits^{(b)} I(U;Z) - H({U_R}) + H({U_R}|{U_I},{u_F},Z)\nonumber\\
&\mathop  \le \limits^{(c)} I(X;Z) - H({U_R}) + H({U_R}|{U_I},{U_F},Z)\nonumber\\
&\mathop  \le \limits^{(d)} NI({W_w}) - H({U_R}) + H({U_R}|{U_I},{U_F},Z)
&
\end{align}
Equation (a) follows from the chain rule of mutual information and a consequence of the following:
\begin{align}
&I(U;Z) = I({U_{{I} \cup {R} \cup {{{{ F}}}}}};Z)\nonumber\\&
 = I({U_I},{U_R},{U_{{{F}}}};Z)\nonumber\\&
 = I({U_{{{ F}}}};Z) + I({U_I},{U_R};Z|{U_{{{ F}}}})\nonumber\\&
 = I({U_I},{U_R};Z|{U_{{{F}}}})&
\end{align}
The last equality in (29) derived from the fact that $I({U_{{{\cal F}_m}}};Z_1^N)$ equals to zero because ${u_{{{\cal F}_m}}}$ has been sent over bad channels for both main and wiretap channels. 
(b) follows form the independence of ${U_R},{U_I},{U_{{{ F}}}}$.(c) is the result of applying the data processing inequality because of the Markov chain $M \to U \to X \to (Y,Z)$. Below operations lead to (d):
\begin{align}
&I(X_1^N;Z_1^N) = H(Z_1^N) - H(Z_1^N|X_1^N)\nonumber\\&
 = H(Z_1^N) - \sum\limits_{i = 1}^N {H({Z_i}|{X_i})} 
 \le \sum {(H({Z_i})}  - H({Z_i}|{X_i})) \nonumber\\&
 = \sum\limits_{i = 1}^N {I({X_i};{Z_i})}  \le NI({W_w})&
\end{align}
According to (d) for finding an upper bound for $I(M;Z|{U_{{{ F}}}})$ an upper bound for $H({U_R}|{U_I},{U_{{{F}}}},Z)$ should be found. For the upper bound of $H({U_R}|{U_I},{U_{{{F}}}},Z)$ we propose the following lemma:

\textbf{Lemma 3}: For a asymmetric binary-input channel the Ensemble of polar code was defined which means defining polar code over all possible choices of frozen symbols. For proposed coding scheme defined over the ensemble there exists a sequence ${u_{{{F}}}}$, which with utilizing it as frozen vector, the following inequality will be satisfied:
\begin{align}
 &H({U_R}|Z,{U_I},{U_F}) \nonumber\\
 &\le H((q - 1){2^{ - {N^\beta }}}) + (r{\log _2}q) \times (q - 1){2^{ - {N^\beta }}}& 
\end{align}
\vspace{-5mm}
\begin{proof}
we define an error event as following:
\begin{eqnarray}
 E = \left\{ {\begin{array}{*{20}{c}}
{\textbf{1          } {{\hat U}_R} \ne {U_R}}\\
{\textbf{0          } {{\hat U}_R} = {U_R}}
\end{array}} \right.
\end{eqnarray}
random vector is sent over good channels for both main channel and wiretap channel so we can write ${P_e} = {P_{B,e}}({{\cal A}_w})$ and:
\begin{align}
&{P_e} = P(E = 1) \nonumber\\& =  \Pr ({{\hat U}_R} \ne {U_R})
 \le \sum\limits_{i \in {{\cal A}_w}} {Z({W_w}_N^{(i)})}  \le {(q-1)*2^{ - {N^\beta }}}
\end{align}
Because coding scheme is defined over all possible choices of ${u_{{{\cal F}_m}}} \in {{{\cal X}}^{|{{{\cal F}}_m}|}}$ and the error probability in average sense is smaller than its upper bound,it means that there exists a specific frozen vector  ${u_{{\cal F}_m}}$ which is in set ${{{\cal X}}^{|{{{\cal F}}_m}|}}$
and with choosing it the error probability does not exceed the upper bound $(q-1)*{2^{ - {N^\beta }}}$.
Now the term $H(E,{U_R}|{U_I},{U_{{{F}}}},Z)$ is expanded in two ways:
\begin{align}
&H(E,{U_R}|{U_I},{U_{{{F}}}},Z) = \nonumber\\&
H({U_R}|{U_I},{U_{{{F}}}},Z) + H(E|{U_R},{U_I},{U_{{{F}}}},Z) =\nonumber\\&
 H({U_R}|E,{U_I},{U_{{{F}}}},Z) + H(E|{U_I},{U_{{{F}}}},Z)&
\end{align}
obviously the term $H(E|{U_R},{U_I},{U_{{{F}}}},Z)$ equals zero because with having $U_R$ there is no equivocation on Error, consequently:
\begin{align}
&H({U_R}|{U_I},{U_{{{F}}}},Z) =\nonumber\\&
H({U_R}|E,{U_I},{U_{{{F}}}},Z) + H(E|{U_I},{U_{{{F}}}},Z)&
\end{align}
clearly eliminating the condition of entropy function do not decrease the entropy so:
\begin{align}
 & H({U_R}|{U_I},{U_{{{F}}}},Z)\nonumber\\& 
 \le H({U_R}|E,{U_I},{U_{{{F}}}},Z) + H(E)&
\end{align}

a new task is to find an upper bound for the right hand side of (36):
\begin{align}
&H({U_R}|E,{U_I},{U_{{{F}}}},Z) = \nonumber\\&
\sum\limits_{i = 0}^1 {P(E = i)H({U_R}|E = i,{U_I},{U_{{{F}}}},Z)}\nonumber\\ &
 = P(E = 1)H({U_R}|E = 1,{U_I},{U_{{{F}}}},Z)
 + (1 - P(E = 1)) \times 0 \nonumber\\&
 = P(E = 1)H({U_R}|E = 1,{U_I},{U_{{{F}}}},Z)\nonumber\\&
 \le P(E = 1)H({U_R}) = {P_e}*r{log_2}q&
\end{align}

according to (36) and (37):
\begin{flalign}
&H({U_R}|{U_I},{U_{{{F}}}},Z) \le H(E) + {P_e}*r{log_2}q\nonumber\\
&\le H((q - 1){2^{ - {N^\beta }}}) + (r{\log _2}q) \times (q - 1){2^{ - {N^\beta }}} &
\end{flalign}
\vspace{-5mm}
\end{proof}
according to lemma 3 and (28):
\begin{align}
&\mathop {\lim }\limits_{N \to \infty } I(M;Z|{U_F})/N = \mathop {\lim }\limits_{N \to \infty } \{ I({W_w}) - r{\log _2}q/N + \nonumber\\
& \{ H((q - 1){2^{ - {N^\beta }}}) + r{\log _2}q*(q - 1){2^{ - {N^\beta }}}\} /N\}&\nonumber\\
&= \mathop {\lim }\limits_{N \to \infty } \{ I({W_w}) - r{\log _2}q/N\}
\nonumber\\
&= {\log _2}q*\mathop {\lim }\limits_{N \to \infty } \{ {I_q}({W_w}) - {H_q}({U_R})\}\nonumber\\
&\approx \mathop {\lim }\limits_{N \to \infty } \{ {I_q}({W_w}) - r/N\}  = 0
\end{align}
${I_q}({W_w})$ is symmetric capacity of wiretap channel if in equation (2), the base of logarithm be q and ${H_q}({U_R})$ is q-based logarithmic entropy function.
If $R$ is the rate of coding then:
\vspace{-3mm}
\begin{align}
\frac{{H(M)}}{N} \le R \le \frac{{k{{\log }_2}q}}{N} &= \frac{{{{\log }_2}|q{|^{k + r}} - {{\log }_2}|q{|^r}}}{N} \nonumber\\
&= I({W_m}) - I({W_w}) = {C_S}
\end{align}

It concludes that for achieving secrecy capacity the distribution of message should be uniform, otherwise there exists a code such that its rate $R$ satisfy the following inequality :

\begin{align}
\frac{{H(M)}}{N} < R < \frac{{k{{\log }_2}q}}{N}
\end{align}
And the code rate doesn't achieve secrecy capacity. Inequality $H(M)/N \le R$ derives from source coding theorem.
So the security of coding scheme is proved because $\mathop {\lim }\limits_{N \to \infty } I(M;Z)/N$ is equivalent to $\mathop {\lim }\limits_{k \to \infty } I(M;Z)/k$. Equation (39) holds because according to the definition of polar coding $I({W_w})\mathop  \equiv \limits^{N \to \infty } |{{\cal A}_w}|/N$.

\section{Conclusion and discussion}
We briefly mention our results and certain extensions of them. So far, nonbinary-input asymmetric wiretap channels have been considered in this paper. We proved that there exists some choices of frozen symbols for which coding scheme satisfies reliability and security conditions. Also it has been proved that the necessary condition for achieving secrecy capacity is uniform distribution of the message set. 
An open problem is to construct codes for the situation where the wiretap channel is not degraded with respect to the main channel.for example if Eve has just a less noisy channel than Bob.
 Our coding scheme is not totally explicit because only existence of a frozen vector that is suitable for our setting has been proven.
Another problem is successive cancellation decoding depends on the past estimates of itself that should be correct otherwise that will propagate error. Implementing another decoder for overcoming this setback is of great interest.
Recently in \cite{hon13} a method has been suggested for overcoming the polarization limitation. The method is based on defining pseudorandom frozen bits and extent to all kind of channels. Using results in \cite{hon13} to extend our results to arbitrary discrete memolyless channels is a thrilling task.
It was shown by Maurer-Wolf \cite{mau20} that any coding scheme that satisﬁes the “weak” security condition can be converted to a coding scheme that satisﬁes the stronger condition (3). This has been accomplished using information reconciliation and privacy ampliﬁcation protocol \cite{ben95}. So the results of this paper could be extented to strong security using results of privacy ampliﬁcation protocol.
\bibliographystyle{elsarticle-harv}

\end{document}